\documentclass[10 pt, twocolumn,secnumarabic,amssymb, nobibnotes, aps, prb,longbibliography,floatfix]{revtex4-2}
\usepackage{float}
\usepackage{amsmath, graphicx} 
\usepackage{float}
\usepackage{tikz}
\usepackage[utf8]{inputenc}
\usepackage[T1]{fontenc}
\usepackage{textcomp}
\usepackage{adjustbox}
\usepackage{silence}
\usepackage[dvipsnames]{xcolor}
\usepackage{braket} 
\usepackage{siunitx} 
\usepackage{svg}
\usepackage[dvipsnames]{xcolor} 
\usepackage{tcolorbox}
\usepackage[
  bookmarks=false,
  breaklinks=true,
  colorlinks=true,
  citecolor=blue,
  linkcolor=blue,
  urlcolor=blue
]{hyperref}

\usepackage{mathabx} 
\usepackage{subfigure}
\usepackage{orcidlink}
\usepackage{microtype}
\usepackage{cleveref}
\usepackage{newunicodechar}
\newunicodechar{̈}{\"{}}
\usepackage{textcomp}
\usepackage{textgreek} 
\usepackage{etoolbox}
\makeatother
\begin{document}

\title{Strain-induced non-trivial band topology and large magnetoresistance in epitaxial \texorpdfstring{SrTaO$_{3}$}{SrTaO3} thin films}

\author{Aditya Mehta\,\orcidlink{0009-0003-2926-6793}$^{1,2}$}
\thanks{\href{mailto:aditya.mehta@iopb.res.in}{aditya.mehta@iopb.res.in}}
\author{Soumyakanta Panda\, \orcidlink{0000-0001-6548-2010}$^{3}$}
\altaffiliation{Present address: Laboratoire Albert Fert, CNRS, Thales, Université Paris-Saclay, 91767 Palaiseau, France}
\author{Pratik Sahu\,\orcidlink{0000-0003-0346-5682}$^{4,5}$}
\author{Amit Chauhan$^{6}$}
\author{Dibyajyoti Sahoo$^{7}$}
\author{Shwetha G Bhat$^{8}$}
\author{Rohit Soni$^{7}$ }
\author{Goutam Tripathy$^{1,2}$}
\author{Niharika Mohapatra\,\orcidlink{0000-0002-3182-8818}$^{3}$}
\author{B. R. K. Nanda\,\orcidlink{0000-0001-8330-4824}$^{4,5}$}
\author{D. Samal\,\orcidlink{0000-0003-2618-4445}$^{1,2}$}
\thanks{\href{mailto:d.samal@iopb.res.in}{dsamal@iopb.res.in}}

\affiliation{$^1$Institure of Physics, Bhubaneswar-751005, India}
\affiliation{$^2$Homi Bhabha National Institute, Anushakti Nagar, Mumbai-400094, India}
\affiliation{$^3$School of Basic Sciences, Indian Institute of Technology Bhubaneswar, Bhubaneswar, Odisha 752050, India}
\affiliation{$^4$Center for Atomistic Modelling and Materials Design, Indian Institute of Technology Madras, Chennai 600036, India }
\affiliation{$^5$Condensed Matter Theory and Computational Lab, Department of Physics, Indian Institute of Technology Madras, Chennai 600036, India }
\affiliation{$^{6}$Peter Gr\"unberg Institute, Forschungszentrum J\"ulich, 52425 J\"ulich, Germany}
\affiliation{$^7$Department of Physics, Indian Institute of Science Education and Research Berhampur, Berhampur-760010, India}
\affiliation{$^8$Department of Physics, Indian Institute of Science Bangalore-560012, India}
\date{\today}

\begin{abstract}
Epitaxial thin films of complex transition metal oxides constitute an ideal platform for realizing correlated topological phases through strain engineering of crystal symmetry and band topology. Here, we investigate the electronic transport properties and the band structure of the 5$d^1$ paramagnetic metal SrTaO$_3$ thin film under epitaxial compressive strain. The film exhibits a large non-saturating quasi-quadratic magnetoresistance (MR) and a nonlinear Hall effect characteristic of two-band transport, together with the field induced re-entrant metallic behavior. The violation of Kohler's scaling further corroborates the scattering from multiple bands. Semiclassical two-band analysis yields carrier densities and mobilities of the order $\sim 10^{21}~\mathrm{cm^{-3}}$ and $\sim 10^{4}~\mathrm{cm^2 V^{-1} s^{-1}}$, respectively. The observed large mobility and unsaturated MR, indicates the presence of a non-trivial band structure. Consistent with this, results from density functional theory calculations and tight binding model studies reveal that the stable structure has an octahedral rotation $a^0a^0c^+$, where the non-symmorphic symmetry ensures the presence of topological nodal lines yielding finite Berry curvature in the vicinity of the Fermi level. Besides the nodal line, trivial electron and hole pockets exist near the Fermi level, contributing to electron transport. These findings establish that epitaxial strain drives SrTaO$_3$ into a nontrivial topological semimetal, a phase inaccessible in the bulk counterpart.

\end{abstract}

\maketitle
\section{\label{sec:one}Introduction}
Topological semimetals (including Dirac, Weyl, and nodal-line types), characterized by symmetry-protected band crossing and unconventional electronic transport behavior such as chiral anomaly~\cite{PhysRevB.99.045143,ong2021experimental,xiong2015evidence}, ultrahigh carrier mobility, and large non-saturating magnetoresistance~\cite{6ycd-k56z,hu2019transport,shekhar2015extremely}, have attracted considerable interest from basic research as well as quantum technologies~\cite{zhai2024rise,yang2025advances,he2022topological,bao2022light,alfieri2023nanomaterials}. While Dirac and Weyl semimetals feature fourfold and twofold band degeneracies at discrete nodal points in momentum space respectively, topological nodal-line semimetals (NLSMs) feature band crossings that extend along a finite dimension in the form of a continuous line or a loop in the momentum space~\cite{PhysRevMaterials.8.L041201}. This unique topology gives rise to nearly flat drumhead surface states and an enhanced density of states (DOS) near the Fermi level~\cite{bian2016topological}, thereby augmenting electron correlation effects~\cite{chang2019realization} and is also expected to host exotic semi-Dirac fermions~\cite{PhysRevX.14.041057}. 
Although extensive studies have been carried out on topological semimetals whose properties are mainly dominated by $s$- and $p$- bands close to the Fermi energy~\cite{liu2014discovery,samal2016molecular,nakamura2020robust,konig2007quantum,bernevig2006quantum,zhang2009topological}, investigations on correlated topological materials derived from $d$-band electrons remain relatively limited~\cite{fujioka2019strong, itoh2016weyl,PhysRevB.106.085139,takiguchi2020quantum,ghosh2023spontaneous, ok2021correlated,PhysRevLett.127.277204,takiguchi2020quantum,takahashi2018anomalous,hsu2021evidence}. The intersection of electronic correlations and nontrivial band topology represents an important research direction, giving rise to emergent quantum phases such as Mott topological insulators~\cite{PhysRevLett.100.156401}, topological Kondo insulators~\cite{TopologicalKondoinsulator,PhysRevLett.104.106408}, and axion insulators~\cite{liu2020robust}. 

In the realm of materials platform, 4$d$/5$d$ based complex transition-metal oxides (TMOs) provide a fertile ground for realizing correlated topological phases that arise from the interplay of strong Coulomb interactions in partially filled $d$ orbitals and sizable spin-orbit coupling. Despite their potential and extensive theoretical predictions~\cite{zhang2019catalogue, frey2020high, PhysRevB.111.195110,PhysRevB.95.035102}, the experimental realization of such phases in bulk TMOs remain challenging, as electronic correlations and lattice distortions often tend to stabilize competing  symmetry breaking electronic phases that overwhelm the subtle topological features.
Although imposing the requisite non-symmorphic crystalline symmetry could, in principle, protect nontrivial band topology~\cite{RevModPhys.88.035005,yang2018symmetry}, achieving precise control over the requisite lattice symmetry in bulk materials is difficult. In contrast, complex oxide thin films and heterostructures leverage epitaxial strain to systematically tune B-O bond lengths, B-O-B bond angles (B = transition metal cation), and octahedral connectivity, thereby enabling precise control of requisite lattice symmetry to realize exotic topological phases~\cite{ok2021correlated,PhysRevMaterials.3.064205,liu2022emergent,PhysRevB.86.235141,PhysRevB.111.155146,song2026symmetry}.

Mohanta~\textit{et al.}~\cite{PhysRevB.104.235121}, and Rosendal~\textit{et al.}~\cite{PhysRevMaterials.7.075002} predicted that time-reversible cubic perovskites such as SrNbO$_3$, SrMoO$_3$, and CaNbO$_3$ transform into a tetragonal phase accompanied by intrinsic octahedral rotations and can host three-dimensional semi-Dirac fermions. The non-trivial Dirac states are protected by non-symmorphic symmetries of the oxygen octahedral rotations and remain robust against spin-orbit coupling (SOC). Notably, among the above materials, SrNbO$_3$ is the only perovskite oxide in which such a topological semimetallic phase has been experimentally realized through strain engineering~\cite{ok2021correlated,PhysRevB.111.155146,song2026symmetry}. Similar to 4$d^1$ SrNbO$_3$, SrTaO$_3$, an isoelectronic 5$d^1$ analog that crystallizes in a cubic perovskite structure~\cite{ali2015theoretical}, exhibits a nearly identical bulk electronic band structure~\cite{PhysRevMaterials.9.015001}. This close similarity makes SrTaO$_3$ an ideal candidate for exploring the emergence of strain-induced nontrivial topological phase.

To realize this, we fabricated strained tetragonal SrTaO$_3$ thin films on (001)-oriented SrTiO$_3$ substrates and carried out detailed magnetotransport measurements. The longitudinal resistivity exhibits an upturn below the characteristic turn-on temperature ($T_{on}$) under high magnetic fields, followed by a reentrant metallic behavior at low temperatures. The magnetoresistance shows no sign of saturation up to 7~T and reaches a large value of 2.17~$\times$~10$^3$~\%, while the Hall response exhibits a nonlinear dependence, indicative of multicarrier transport. Analysis using a two-band model yields high carrier mobilities on the order of 10$^4$~cm$^2$~V$^{-1}$~s$^{-1}$ and carrier concentrations approximately $10^{21}$ cm$^{-3}$. The violation of Kohler’s scaling law suggests the presence of multiple scattering relaxation processes originating from distinct bands near the Fermi level. Density functional calculations supplemented with Wannier tight-binding led Berry curvature calculations establish the formation of a nodal line in the ground-state electronic structure of the strained tetragonal phase of SrTaO$_3$, which is characterized by the $a^0a^0c^+$ octahedral rotation pattern. Besides the nodal line, trivial electron and hole pockets are also present near the Fermi level.

\section{\label{sec:exp}Experimental section}
\subsection{\label{sec:A}Thin film fabrication}
SrTaO$_{3}$ thin films were fabricated \textit{via} pulsed laser deposition (PLD) on (001) oriented SrTiO$_{3}$ using an in-house synthesized polycrystalline ceramic Sr$_2$Ta$_2$O$_7$ target prepared using a conventional solid-state reaction route~\cite{PhysRevB.108.235114}. 
The substrate temperature was ramped from room temperature to 620$^\circ$C at a rate of 10$^\circ$C/min and maintained during the deposition. The target was then ablated using a KrF excimer laser (248 nm) operating at a repetition rate of 3 Hz with a laser fluence of 0.95 J/cm$^2$, under a partial oxygen pressure of 10$^{-7}$ mbar.
Subsequently, this sintered target was ablated using a KrF excimer laser (248 nm) at a repetition rate of 3~Hz using a fluence of 0.95 J/cm$^2$, under a partial oxygen pressure P(O$_2$) of 10$^{-7}$ mbar. 
Later, thin films were cooled to room temperature at a rate of 10$^\circ$C/min under the same P(O$_2$) condition.

\subsection{\label{sec:B}X-ray diffraction and Transport}
The structural characterization of the thin films was carried out using \textit{ex-situ} Rigaku SmartLab high-resolution X-ray diffraction (HR-XRD) with a Cu K$_\alpha${$_1$} radiation source ($\lambda = 1.54059$~\AA). The electrical transport properties were measured using a Quantum Design Physical Property Measurement System (PPMS), equipped with a 7~T magnetic field, over a temperature range of 2--300~K.

\subsection{\label{sec:C} DFT Calculation methodology} 

The Density functional theory (DFT) calculations are performed on bulk and compressively strained SrTaO$_3$ structure to find the most stable Glazer tilt pattern and its associated electronic structure. To achieve it, we used the projector augmented wave (PAW) \cite{Kresse1999} method as implemented in the Vienna ab-initio simulation package (VASP) \cite{Kresse1996}. The 2 × 2 × 2 supercells were used to create different Glazer tilt patterns. The generalized gradient approximation (GGA) was adopted for the exchange-correlation functional. The effect of correlation was incorporated via the effective interaction parameter $U_{\rm eff} = U - J$ through the rotationally invariant Dudarev’s approach \cite{Dudarev1998}. We used $U_{\rm eff}$ = 2 eV, which is typically the onsite Coulomb interaction strength for strongly spin-orbit coupled systems. The Brillouin-zone integration was carried out using a $6 \times 6 \times 6$ k-mesh to achieve self-consistency. The PAW basis functions include 4$s$, 4$p$, and 5$s$ orbitals for Sr; 2$s$ and 2$p$ orbitals for O; and 5$d$ and 6$s$ orbitals for Ta. The Berry curvature, which was needed to distinguish the topological behavior of the bands, is calculated in the following way. A self-consistent DFT calculation was performed using Quantum ESPRESSO \cite{Giannozzi_2009} package with 10 × 10 × 12 Monkhorst-Pack k-point mesh using norm-conserving PBE \cite{PBE1996} exchange-correlation functionals. The bandstructure from DFT is used as an input to calculate the maximally localized Wannier function (MLWF) using Wannier90 \cite{Pizzi_2020}. Employing the MLWF, a tight-binding model was developed, and the latter was used to estimate the Berry curvature.

\section{\label{sec:Two}Results and Discussion}
\subsection{\label{sec:1}Structural characterization}
\begin{figure}[htbp]
    \centering 
    \includegraphics[scale=0.375]{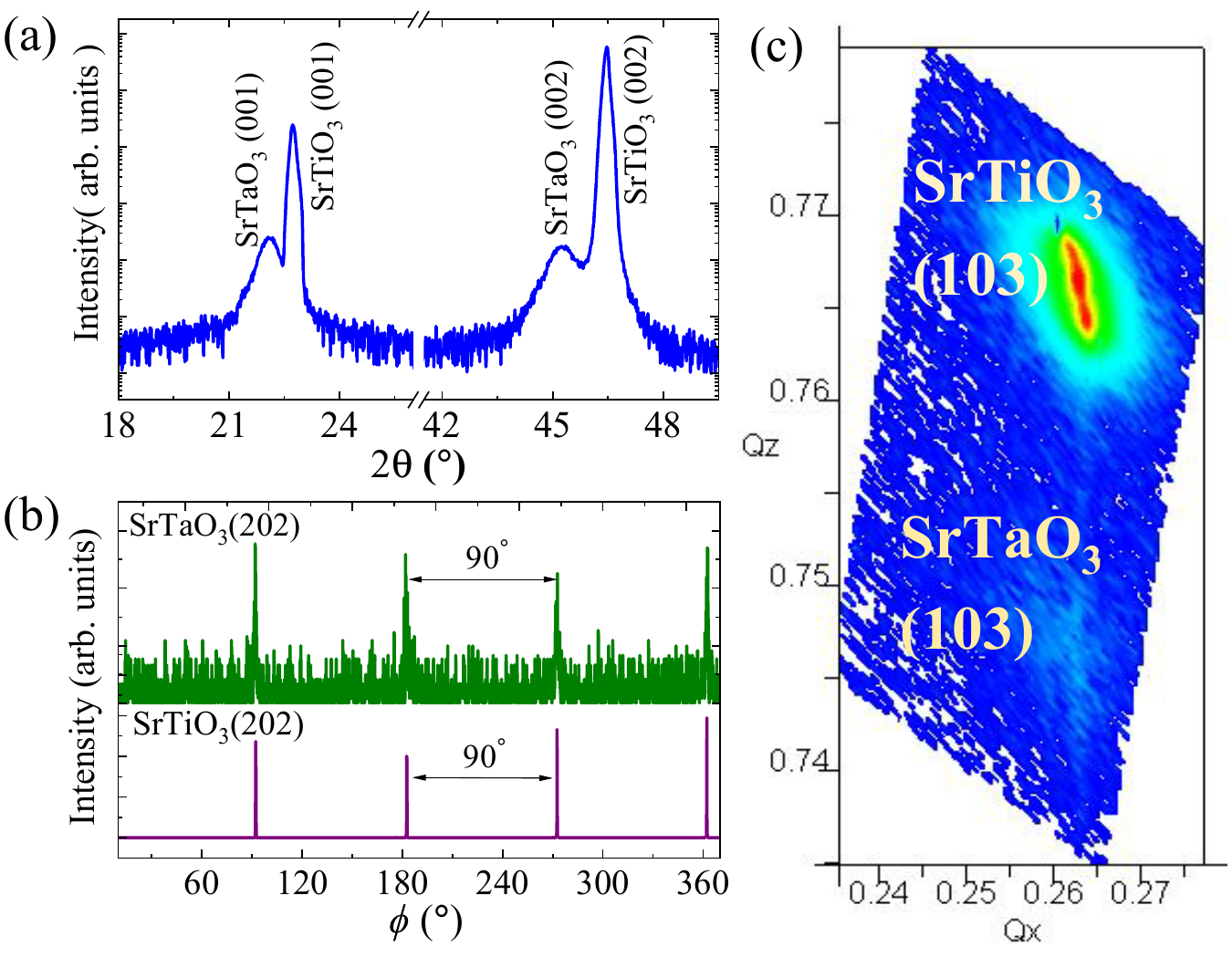}
    \caption {(a) X-ray diffraction (XRD) pattern of SrTaO$_{3}$ thin films grown on (001)-oriented SrTiO$_{3}$ substrate. (b) The $\phi$-scan performed about the (202) reflection of SrTaO$_3$ and SrTiO$_3$ exhibits clear four-fold symmetry, confirming the epitaxial relationship. (c) Reciprocal space mapping (RSM) around the (103) reflection indicates that the film is coherently strained.  }. 
\label{fig: FIG. 1.}
\end{figure}
HR-XRD measurements on SrTaO$_{3}$ thin films confirmed a highly oriented growth of the film on SrTiO$_{3}$ (001). Fig.~\ref{fig: FIG. 1.}(a) exhibits the $\theta$-2$\theta$ scan of SrTaO$_{3}$ film of $\sim$19 nm with $(00l)$ orientation on SrTiO$_{3}$.
 Film thickness was estimated from X-ray reflectivity (XRR) measurement, which was fitted using GlobalFit software (as shown in Supplementary Information Fig. S1(a)). 
Additionally, the $\phi$-scan performed around the (202) reflection, as shown in Fig.~\ref{fig: FIG. 1.}(b), reveals a four-fold symmetry for both the substrate and the film, confirming the epitaxial growth of the film.

Bulk SrTaO$_3$ crystallizes in a cubic perovskite structure with a lattice constant of a = 4.007  \AA{}~\cite{ali2015theoretical}.  Hence, a thin film of SrTaO$_3$ grown on a SrTiO$_3$ (a = 3.905~\AA) substrate exhibits compressive growth with a 2.6$\%$ lattice mismatch relative to SrTiO$_3$. To determine the nature of the strain in the film with respect to the substrate, we performed a reciprocal space mapping (RSM) of the (103) reflection of the film, and the data are shown in Fig.~\ref{fig: FIG. 1.}(c). A highly strained growth of SrTaO$_3$ is observed with respect to SrTiO$_3$, and the estimated out-of-plane lattice parameter is $\sim$4.02~\AA, which is consistent with the literature~\cite{zhang2022design}.

\subsection{\label{sec:2}Temperature dependent resistivity and magnetoresistance}
\begin{figure*}[htbp]
\centering 
\includegraphics[scale=0.265]{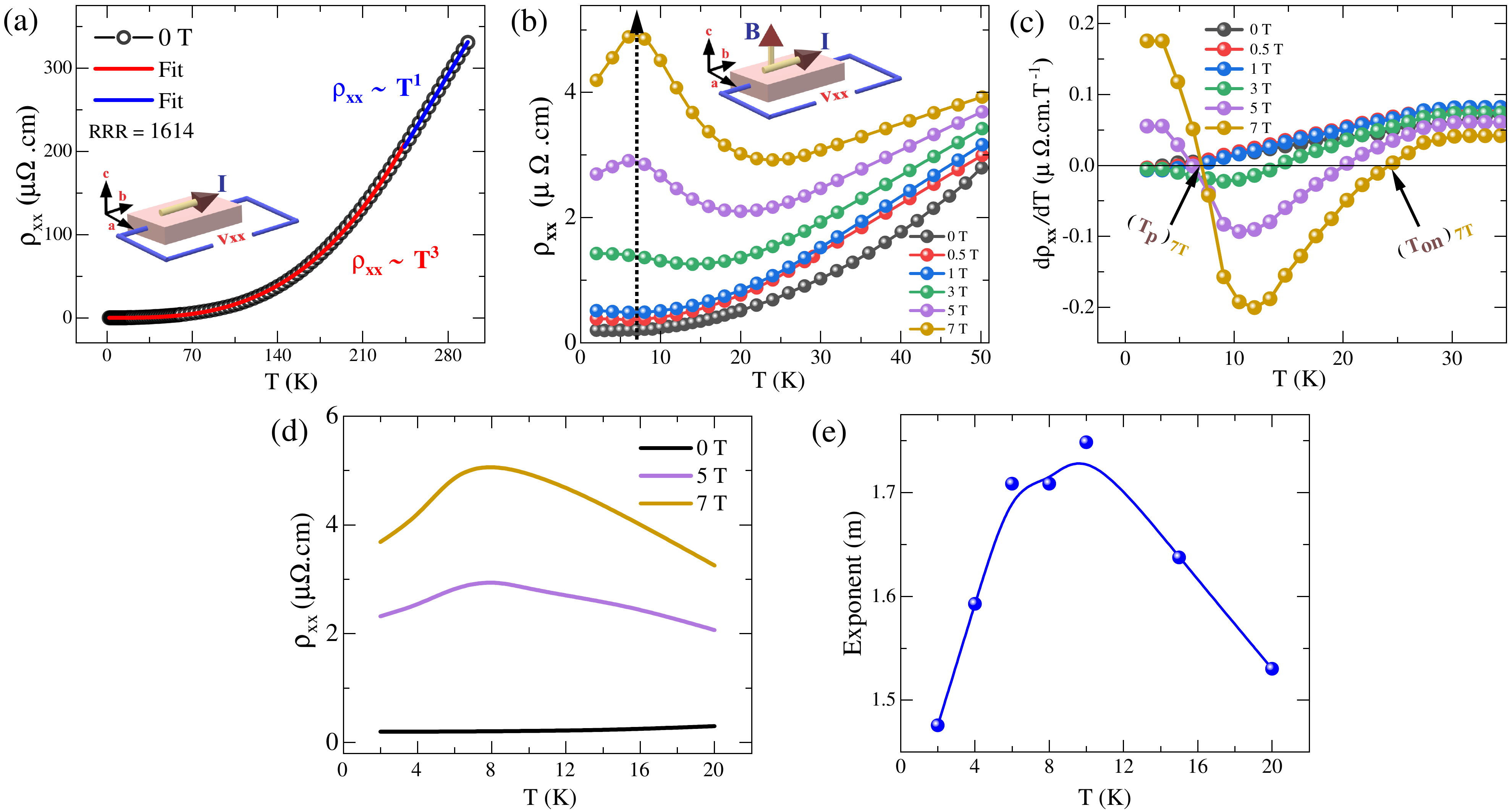}
\caption{ (a) Temperature-dependent zero-field resistivity curve of SrTaO$_3$ films from 300 K to 2 K. (b) Temperature-dependent RT curves from 50 K to 2 K were performed under the application of an external magnetic field. (c) Derivative curve of temperature dependent longitudinal resistivity at various fields.  (d) Simulated longitudinal resistivity $\rho_{xx}(T < 20~\mathrm{K}, \mu_{0}H)$ at various magnetic field strengths (0, 5, and 7~T), illustrating the field dependent evolution of the resistive peak arising from the variation of the exponent $m$ extracted from the magnetoresistance curves shown in (e).} 
\label{fig: FIG. 2.}
\end{figure*}
Fig.~\ref{fig: FIG. 2.}(a) displays the temperature dependent zero-field longitudinal resistivity, $\rho_{xx}(T, 0)$, of the SrTaO$_3$ thin film, which remains metallic over the entire temperature range from 300~K [$\rho_{xx}(300~\mathrm{K}) \sim 331~\mu\Omega\cdot\text{cm}$] down
 to 2~K [$\rho_{xx}(2~\mathrm{K}) \sim 0.2~\mu\Omega\cdot\text{cm}$], resulting in a high residual resistivity ratio (RRR) $[\rho(300~\mathrm{K})/\rho (2~\mathrm{K})]$ of $\sim$ 1614. The observed high RRR indicates the excellent metallicity and high crystallinity of the film.
For temperatures below 245~K ($T < 245~\mathrm{K}$), the zero-field resistivity $\rho_{xx}(T, 0)$ fits well to the expression $\rho_{xx}(T) = \rho_0 + \mathrm{A} T^{n}$, where $\rho_0$ is the residual resistivity at 2~K, $\mathrm{A}$ is a weighting
parameter, and $n$ is a scattering-dependent parameter~\cite{PhysRevB.108.235155}.
The fitting yields $A = 1.25 \times 10^{-5}~\mathrm{\mu}\Omega\cdot\mathrm{cm}\cdot\mathrm{K}^{-3}$ and an exponent $n \sim 3$, suggesting that the $\rho_{xx} (T,0)$ is governed by interband electron–phonon scattering in this temperature regime. This behavior is consistent with that reported for non-trivial topological semimetals such as ZrSiS~\cite{singha2017large}, LaBi~\cite{sun2016large}, RuO$_2$~\cite{peng2025universal}, and NbP~\cite{PhysRevB.96.115152}. 
However, at higher temperatures ($T \geq 245$~K), the resistivity $\rho_{xx} (T,0)$ exhibits a linear temperature dependence, suggesting the $\rho_{xx} (T,0)$ is governed by intraband electron-phonon scattering, similar to that observed in ZrSiS~\cite{singha2017large} and LaBi~\cite{sun2016large}.\\

Figs.~\ref{fig: FIG. 2.}(b) and~\ref{fig: FIG. 2.}(c) display the temperature and field dependent longitudinal resistivity, $\rho_{xx}(T,\mu_{0}H)$, and its derivative, $d\rho_{xx}/dT$, measured under various magnetic field strengths applied along the crystallographic $c$-axis, perpendicular to the current direction. For fields below 1~T ($\mu_{0}H < 1$~T), $\rho_{xx}(T)$ decreases monotonically upon cooling, 
and the positive derivative $(d\rho_{xx}/dT>0)$ indicates that the metallic behavior persists down to 2~K.  
However, when the applied magnetic field is above 1~T, a qualitatively distinct transport regime emerges. 

The resistivity $\rho_{xx}(T, \mu_{0}H > 1~\mathrm{T})$ exhibits an upturn below the characteristic turn-on temperature ($T_\mathrm{on}$, i.e., $T < 25~\mathrm{K}$), evident from a negative slope $(d\rho_{xx}/dT < 0)$ [Fig.~\ref{fig: FIG. 2.}(c)].
To elucidate the origin of the turn-on behavior, we consider the field dependent longitudinal resistivity within the framework of the semiclassical two-band model as
\begin{equation}
\rho_{xx}(T,\mu_{0}H) = \rho_{xx}(T,0) + \frac{\alpha\, H^{m}}{\rho_{xx}^{\,m-1}(T,0) }
\label{eq:rho_xx_field}
\end{equation}
where $\rho_{xx}(T,0)=1/[ e (n_{e}\mu_{e} + n_{h} \mu_{h})]$, and $n$, $\mu_{e}$, and $\mu_{h}$ represent the temperature dependent carrier density, electron mobility, and hole mobility, respectively.
Because of the high mobility ($\sim 10^{4}~\mathrm{cm^2 V^{-1}} s^{-1}$) at low temperatures (discussed in Sec.~\ref{sec:3}) and its reciprocal relation with $\rho_{xx}(T,0)$, the first term becomes negligible at lower temperature, while the second term dominates, leading to an upturn below the turn-on temperature. Similar turn-on behavior has been observed in other materials that exhibit large, non-saturating magnetoresistance, such as WTe$_2$~\cite{PhysRevB.92.180402}, NbP~\cite{PhysRevB.96.115152}, and ZrSiS~\cite{singha2017large}.
\begin{figure}[htbp]
\centering 
\includegraphics[scale=0.30]{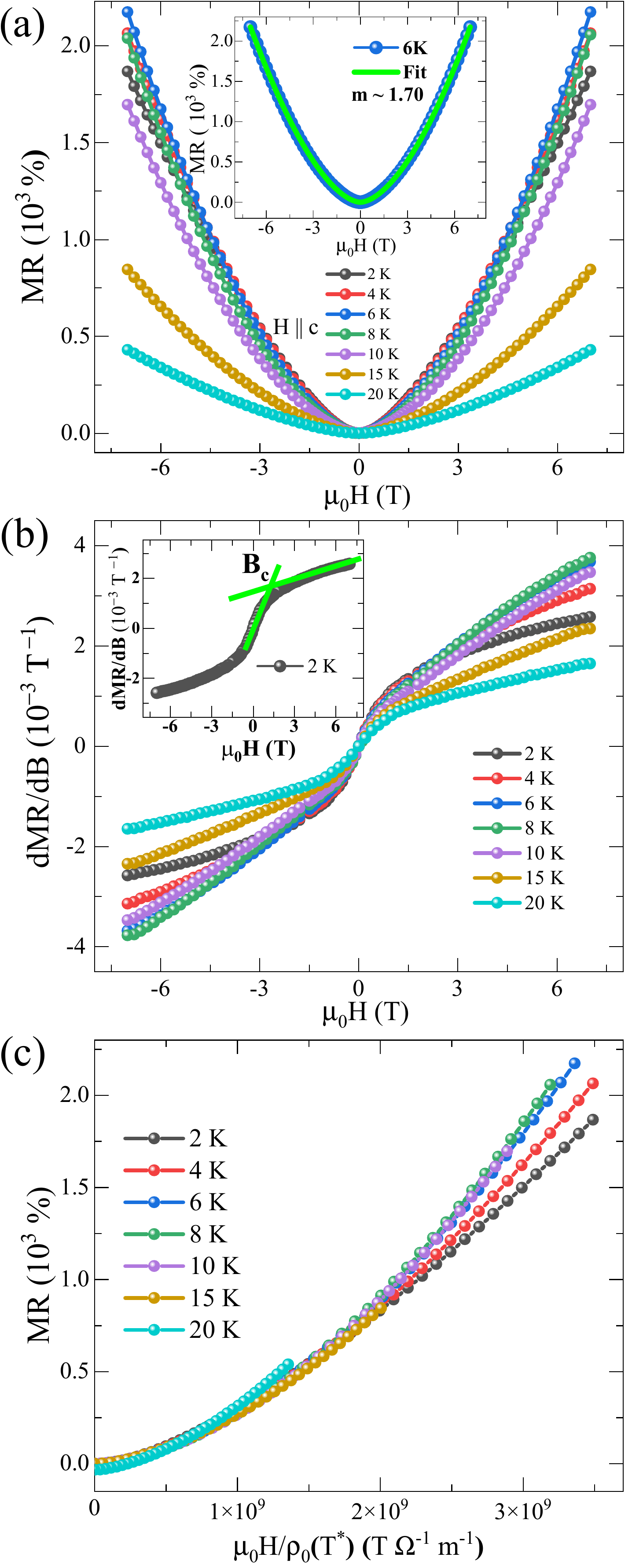}
\caption{(a) Field-dependentt magnetoresistance curves at different temperatures. (b) Derivative of magnetoresistance as a function of magnetic field at different temperatures. Inset: The critical magnetic field extraction from the derivative of MR plots. (c) Conventional Kohler's plot of magnetoresistance.}
\label{fig: FIG. 2_1.}
\end{figure}
Interestingly, for temperatures below 8~K ($T < 8$~K) and under the application of magnetic fields above 3~T ($\mu_{0}H > 3$~T), the metallic character reemerges $(d\rho_{xx}/dT>0)$ at the onset of temperature $T_\mathrm{p}$. This reentrant metallic behavior closely resembles that observed in several topologically trivial and nontrivial semimetals, including NbP~\cite{PhysRevB.96.115152}, TaP~\cite{doi:10.1073/pnas.1808747115}, and graphite~\cite{ PhysRevLett.90.156402,konenkova2008metal, PhysRevB.73.165128,PhysRevLett.90.156402}. Recently Zhang~\textit{et al.}~\cite{zhang2025inadequacy} proposed a unified framework to account for such behavior by considering temperature and field induced evolution of Fermi surface topology that govern the underlying scattering rate and hence shape the $\rho_{xx}(T)$ curve in a magnetic field.
To gain qualitative insight into the origin of the observed reentrant behavior in the present case, we performed numerical simulations of the longitudinal resistivity $\rho_{xx}(T,\mu_{0}H)$ using equation~\ref{eq:rho_xx_field} and which can be rewritten as  
\begin{equation}
\rho_{xx}(T,\mu_{0}H) = \rho_{0}+ A~T^{n}+ \frac{\alpha\, H^{m}}{(\rho_{0}+A {T}^{n})^{m-1} }
\label{eq:rho_xx_1_field}
\end{equation}

For these simulations, we used the parameters extracted from the temperature dependent longitudinal resistivity $\rho_{xx}(\mathrm{T\leq 20~K}, \mu_{0}\mathrm{H} = 0- 7~\mathrm{T})$ together with the temperature dependent exponent $m$ (Fig.~\ref{fig: FIG. 2.}(f)), which was obtained from the magnetoresistance power-law analysis (discussed later in this section). The simulated curves capture the reentrant characteristics observed in the experimental data (Fig.~\ref{fig: FIG. 2.}(e)). Notably, when $m$ is held fixed as a function of temperature, the reentrant behavior does not appear. This suggests that the observed reentrant behavior may originate from a Fermi-surface reconstruction at low temperatures under the application of magnetic fields, which is further supported by the imbalance in charge-carrier densities discussed in Section~\ref{sec:3}.

\begin{figure*}[htbp]
\centering 
\includegraphics[scale=0.22]{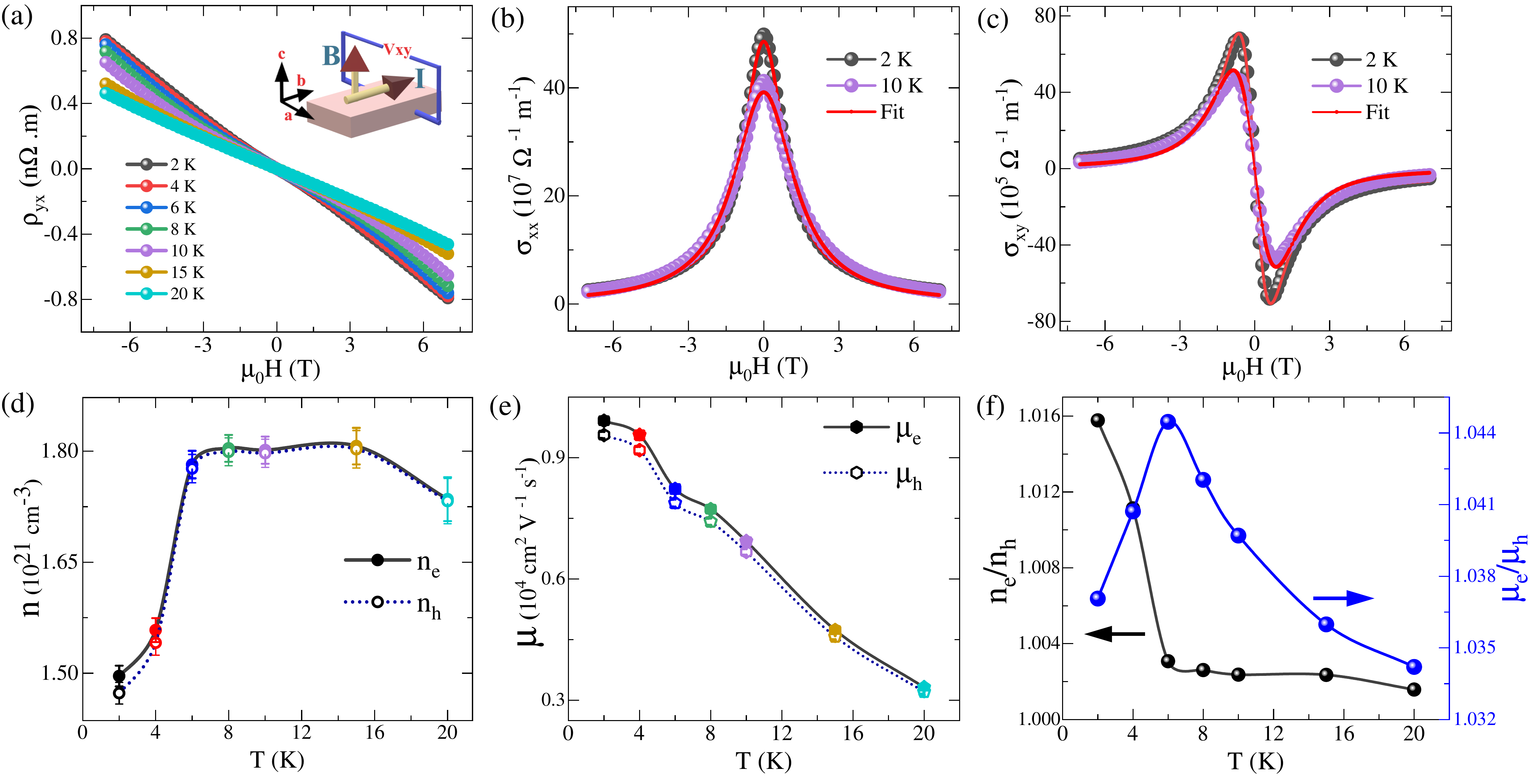}
\caption{(a) Hall measurement performed when H||c at different temperatures. (b-c) Longitudinal and transverse conductivities at different temperatures fitted with a two-band model. (d-e) Temperature dependent charge carrier densities and mobilities extracted from the two-band model fitting. (f) The variation of charge carrier density and mobility ratios with temperature.}
\label{fig: FIG. 3.}
\end{figure*}
In Fig.~\ref{fig: FIG. 2_1.}(a), we plot the magnetoresistance (MR) at different temperatures under the applied magnetic field perpendicular to the current direction. The MR was calculated using the following equation: 
\begin{equation}
\mathrm{MR}~(T, \mu_0H) = \frac{\rho_{xx}~(T, \mu_0H) - \rho_{xx}~(T, 0)}{\rho_{xx}(T, 0)}
\end{equation}
where \(\rho_{xx}~(T, \mu_0H)\) is the longitudinal resistivity in the presence of a magnetic field and \(\rho_{xx}~(T, 0)\) is the zero-field longitudinal resistivity.

At 2~K, the MR reaches a notably high value of $1.87 \times 10^{3}$\% and shows no sign of saturation up to 7~T, which is comparable to that observed in several non-trivial topological semimetals~\cite{PhysRevB.94.041103,yanda2025direct}. Intriguingly, the MR increases with temperature and attains to a maximum value of $\sim 2.17 \times 10^{3}$\% at 6~K, which is about 16.2$\%$ larger than the value observed at 2~K. This unconventional enhancement of MR at 6~K, relative to 2~K, can be ascribed to the temperature evolution of charge carrier compensation and mobilities, as discussed in the section~\ref{sec:3}. However, for temperatures above 6~K ($T > 6$~K), the MR decreases monotonically with increasing temperature, reaching approximately $4.30\times 10^{2}$\% at 20~K.

Typically, MR follows a power-law (\(\mathrm{MR}~\propto~\mathrm{H}^{m}\)) dependence on the magnetic field ~\cite{chen2018large, PhysRevX.11.041029}. The maximum MR measured at 6~K exhibits a distinct power-law scaling with an exponent of \(m \approx 1.70\), indicating a quasi-quadratic field dependence, as illustrated in the inset of Fig.~\ref{fig: FIG. 2_1.}(a).
To further examine this behavior, we analyzed the derivative of MR with respect to the magnetic field, as shown in Fig.~\ref{fig: FIG. 2_1.}(b). The inset of Fig.~\ref{fig: FIG. 2_1.}(b), shows the representative $d\mathrm{MR}/dB$ plot at 2~K with characteristic critical magnetic field ($B_c$) of $\sim$1.24~T. Below $B_c$, $d\mathrm{MR}/dB$ varies linearly with the magnetic field, indicating the quadratic dependence of MR on the applied field. However, beyond $B_c$, $d\mathrm{MR}/dB$ deviates from linearity and exhibits a non-linear field dependence, confirming the quasi-quadratic nature of MR at higher field strengths.

This quasi-quadratic behavior points to the presence of nearly compensated electron and hole charge carriers~\cite{chen2020large}, which is corroborated by the Hall measurement discussed in Section~\ref{sec:3}. The presence of nearly compensated charge carriers hints at the possible existence of multiple Fermi surface (FS) pockets near the Fermi level. This is further examined using Kohler’s scaling law~\cite{PhysRevX.11.041029, PhysRevB.109.035129}, which holds only when a single scattering relaxation time governs the scattering process over the entire FS.
However, in the present case, a clear violation of Kohler’s scaling law is observed, as the MR curves at different temperatures as a function of ${\mu_0\mathrm{H}}/\rho_0(T^{*})$ do not collapse onto a single curve (Fig.~\ref{fig: FIG. 2_1.}(c)), indicating the presence of multiple scattering times originating from the multiple FSs~\cite{yanda2025direct},  
which leads to large non-saturating MR. Such observation has been made in the nodal-line semimetal MoO$_2$~\cite{chen2020large} and the Dirac semimetal BiRe$_2$O$_6$~\cite{yanda2025direct}.

\subsection{\label{sec:3}Hall Effect Measurements}

Figure~\ref{fig: FIG. 3.}(a) shows the Hall measurements performed at different temperatures. The measured Hall resistivity exhibits nonlinearity with the magnetic field, which is indicative of multicarrier transport~\cite{PhysRevB.111.144402, ye2025extremely}. To determine the charge carrier densities and corresponding mobilities, the longitudinal and Hall resistivity data were analyzed using a semiclassical two-band model. In this model, the conductivity tensor, in its complex form, is given by~\cite{chen2020large,ali2014large, PhysRevB.96.115152}
\begin{equation}
\sigma(H) = \frac{e~ n_e \mu_e}{1 + i \mu_e \mu_0 H} + \frac{e~ n_h  \mu_h}{1 - i \mu_h \mu_0 H}
\label{eq.4}
\end{equation}

where $n_e$ and $n_h$ are the electron and hole carrier densities, respectively, and $\mu_e$ and $\mu_h$ are their respective mobilities.
The real and imaginary components of this tensor correspond to the longitudinal and transverse conductivities, respectively, and are given by
\begin{equation}
\sigma_{xx} = e \left[ \frac{n_h \mu_h}{1 + (\mu_h \mu_0 H)^2} + \frac{n_e \mu_e}{1 + (\mu_e \mu_0 H)^2} \right]
\label{eq.5}
\end{equation}
\begin{equation}
    \sigma_{xy} = e \mu_0 H \left[ \frac{n_h \mu_h^2}{1 + (\mu_h \mu_0 H)^2} - \frac{n_e \mu_e^2}{1 + (\mu_e \mu_0 H)^2} \right]
    \label{eq.6}
\end{equation}

To apply this model to the experimental data, we converted the measured longitudinal resistivity $\rho_{xx}(T, \mu_0H)$ and the Hall resistivity $\rho_{yx}(T, \mu_0H)$ into conductivities using the standard tensor relations such as
$\sigma_{xx} = {\rho_{xx}}/{(\rho_{xx}^2 + \rho_{yx}^2)}$
and $\sigma_{xy} = {\rho_{yx}}/{(\rho_{xx}^2 + \rho_{yx}^2)}$~\cite{chen2020large,ali2014large,PhysRevB.96.115152}, respectively.
The extracted conductivities were then analyzed through simultaneous fitting using the real and imaginary parts of the conductivity tensor~\eqref{eq.4} to estimate the charge carrier densities and mobilities at different temperatures. 
The fits show good agreement with the experimental data, as illustrated in Figs.~\ref{fig: FIG. 3.}(b) and \ref{fig: FIG. 3.}(c) for 2 K and 10 K and fitting results for the intermediate temperatures (4 K, 6 K, and 8 K) are provided in the Supplementary Information (Figs.~S3(a–f)).

The extracted temperature dependent charge carrier densities and their respective mobilities obtained from the simultaneous fitting are displayed in Fig.~\ref {fig: FIG. 3.}(d-e). The charge-carrier densities are of the order of $10^{21}\mathrm{cm}^{-3}$, comparable to those values reported for topological nodal-line semimetals such as ZrB$_2$~\cite{PhysRevB.97.205105}, ZrSiSe~\cite{PhysRevLett.117.016602}, ZrSiTe~\cite{PhysRevLett.117.016602}, and MoO$_2$~\cite{chen2020large}.
At 2~K, charge carrier mobilities reach approximately $10^{4} \mathrm{cm}^{2} \mathrm{V}^{-1}\mathrm{s}^{-1}$ and decrease monotonically with increasing temperature, as shown in Fig.~\ref{fig: FIG. 3.}(e), which can be attributed to phonon thermal scattering at higher temperatures~\cite{chen2020large}. As shown in Fig.~\ref{fig: FIG. 3.}(f), at 2~K the charge carrier density ratio ($n_{e}/n_{h}$) is 1.016, indicating slight electron domination, which is 1.6\% larger than the hole density and hence partial compensation of charge carriers. With increasing temperature, the ratio approaches unity and reaches nearly perfect compensation at 6~K ($n_{e}/n_{h} \approx 1.003$), which reflects in a maximum MR,  consistent with the temperature dependence of the mobility ratio $\mu_{e}/\mu_{h}$ shown in Fig.~\ref{fig: FIG. 3.}(f). For $T > 6$~K, the ratio of carrier densities remains close to unity ($n_{e}/n_{h} \approx 1.004$). However, the carrier mobilities decrease with increasing temperature, leading to a gradual reduction in MR.

\subsection{\label{sec:4} Insights from electronic structure studies}
\begin{figure*}[htbp]
    \centering 
    \includegraphics[width=\textwidth]{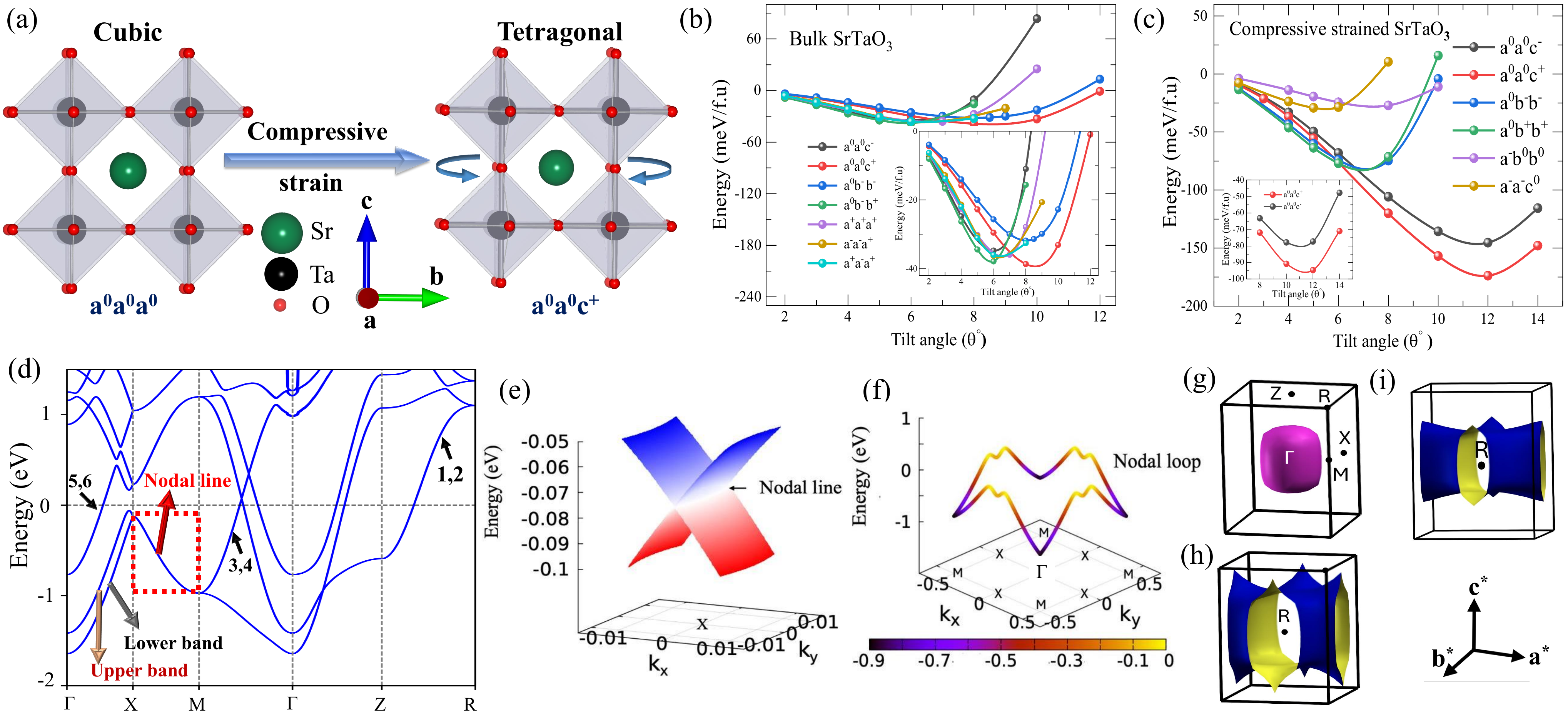}
        \caption {(a) Strain induced cubic-to-tetragonal phase transition with in-phase octahedral rotation along $c$-axis.  (b,c) The energy optimization curves as a function of tilting angle of various Glazer tilting patterns for bulk and compressively strained structures ($\simeq 2.6\%$ in
the ab plane, which is the case when SrTaO$_3$ is grown on SrTiO$_3$ substrate). The energy of the untilted structure is set to zero. In both cases $a^0a^0c^+$ forms the ground state with $\theta$ optimized $\approx$ 8$^\circ$ and 12$^\circ$, respectively. The inset shows the result for the thin-film case where 8 formula units of SrTaO$_3$ are grown on 4 formula units of SrTiO$_3$ substrate with a vacuum of 15 \AA. Despite the $a^0a^0c^-$ octahedral distortion in the SrTiO$_3$ substrate, the ground-state configuration of the SrTaO$_3$ film prefers the $a^0a^0c^+$ octahedral rotation pattern.(d) The band structure of the strained tetragonal $a^0a^0c^+$ structure. A nodal line emerges for the case of $a^0a^0c^+$,  along the X-M k-path. (e) Further illustration of the nodal line by plotting the band structure in the 2D momentum space with X point at its center.  (f) The loop formed by the nodal line forms in the full Brillouin zone.(g) Electron pocket at $\Gamma$, formed by bands 5 and 6, and (h) hole pocket at R, formed by band 1 and 2 for E =$\varepsilon_F$ and (i) for E = $\varepsilon_F$ + 0.5 eV.}
\label{fig: Fig. 0.}
\end{figure*}
The origin of experimentally observed non-trivial transport is elucidated by carrying out DFT calculations followed by the Berry curvature analysis with the aid of Wannier tight-binding calculations. The calculations are performed on bulk and compressively strained SrTaO$_3$ structures with different Glazer tilt patterns. As experimentally observed, we have considered $\approx$ 2.6\% compressive strain, which arise due to the lattice mismatch between SrTaO$_3$ and the substrate SrTiO$_3$. \\ 

The Figure~\ref{fig: Fig. 0.}(a) illustrates the undistorted cubic $a^0a^0a^0$ and compressively strained tetragonal $a^0a^0c^+$ SrTaO$_3$ crystal structures. 
The Figs.~\ref{fig: Fig. 0.}(b) and~\ref{fig: Fig. 0.}(c) shows the energetics of different glazer patterns (see SM Fig. S3) as a function of tilt angle ($\theta$) for bulk and strained structures. In these calculations, the angle of rotation along the crystal axes $a$, $b$ or $c$ is kept same. 
As inferred from Fig.~\ref{fig: Fig. 0.}(b), the $a^0a^0c^+$ structure (in-phase rotation about the $c$ axis) gains energy with respect to the untilted structure and is most stable with $\theta \approx 8^\circ$. However, upon close inspection, we also note that this ground state is energetically close (with in the range of $\approx$ 4 meV) to a few other configurations emerging out of the optimization of  other tilting patterns such as $a^+a^+a^+$, $a^0b^-b^+$, etc. With compressive strain, $a^0a^0c^+$ ground state remains robust forming an energy minimum at $\theta \approx 12^\circ$. Contrary to the unstrained case, at the minimum, $a^0a^0c^+$ structure has an enhanced energy gain as compared to other tilting systems, except $a^0a^0c^-$ (out-of-phase rotation along c-axis), whose minimum is close to the $a^0a^0c^+$  structure.

The DFT+U+SOC ($U_{eff}$=2) band structure corresponding to the above mentioned ground state $a^0a^0c^+$ structure is shown in Fig.~\ref{fig: Fig. 0.}(d). There exists a line node between high-symmetric k-points X and M both in the valence and conduction band spectrum, which is further illustrated through the 3$d$ surface plots in Fig.~\ref{fig: Fig. 0.}(e), centered around the X point. Since there are four X and M points in the Brillouin zone for the plane $k_z=0$, the nodal line forms a loop, which is shown in Fig.~\ref{fig: Fig. 0.}(f). Consequently, tetragonal SrTaO$_3$ is a nodal line semimetal. Further analysis reveals that the nodal line is protected by the non-symmorphic symmetry of the tetragonal P4/mbm space group~\cite{PhysRevMaterials.7.075002}.\\

\begin{figure}[htbp]
    \centering
    \includegraphics[width=0.8\linewidth]{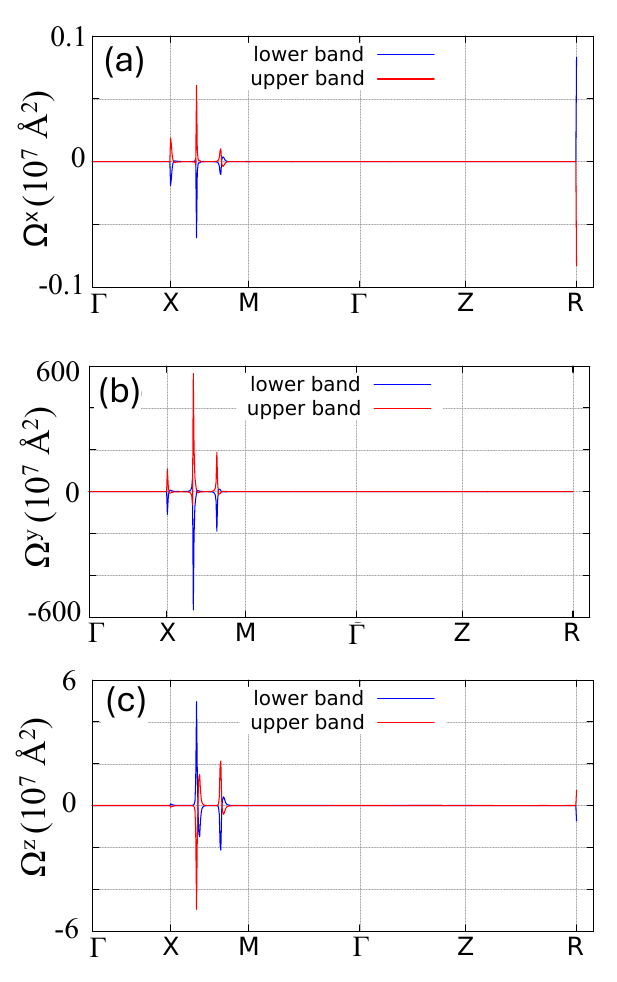}
    \caption{The Berry curvature components for the upper and lower band are plotted for (a) $\Omega^x$, (b) $\Omega^y$, and (c) $\Omega^z$ along the high-symmetry k-path.}
    \label{fig:berry}
\end{figure}

To further validate the presence of nodal line near the Fermi level, we have computed the Berry curvature for the upper and the lower band corresponding to the valence band spectrum (see Fig.~\ref{fig: Fig. 0.}(d)).
The Berry curvature components for any band $n$ are calculated by evaluating the following expression~\cite{10.1098/rspa.1984.0023}.  

\begin{equation}
\Omega_n^{\alpha}(\mathbf{k}) = -2\ Im\sum_{m \ne n} 
\frac{ 
\braket{ u_{n\mathbf{k}} | \partial_{k_\beta} H | u_{m\mathbf{k}}}
\braket{ u_{m\mathbf{k}} | \partial_{k_\gamma} H | u_{n\mathbf{k}}}
}{
(\varepsilon_{m\mathbf{k}} - \varepsilon_{n\mathbf{k}})^2
},
\label{berry}
\end{equation}

where $\alpha$, $\beta$, and $\gamma$ form a cyclic permutation of cartesian indices, H is the Hamiltonian, and $\ket u_{nk}$ is the $n^{th} $ eigenstate. 
We observe sharp Berry curvature peaks between high-symmetric k-points X and M in the Brillouin zone, as shown in Fig.~\ref{fig:berry}, which confirms the topological nature of these bands~\cite{PhysRevB.106.L081101}. Additionally, the exact opposite values for the lower and the upper bands proves the presence of nodal line along X-M, as the TR symmetry is intact~\cite{RevModPhys.82.1959}. We also observe that the magnitude of the Berry curvature component $\Omega^y$ is two-to-three orders higher as compared to $\Omega^x$ and $\Omega^z$. The finite Berry curvature observed at the R point corresponds to the bands lying $\sim$ 1 eV above the Fermi level, and do not have any contribution to the experimental observations made in this work. Additionally, though it appears that there are Dirac-like crossings between high-symmetric k-points M and $\Gamma$, upon close inspection, they turn out to be gapped, which is also reflected in the zero Berry curvature obtained along the $\Gamma$-M path (see Fig.~\ref{fig:berry}). Furthermore, as the band structure of Fig.~\ref{fig: Fig. 0.}(a) indicates, there also exists a nodal line mirroring the valence band nodal line.

In addition to the feature of the nodal line semimetal, the electron and hole pockets formed by the electronic structure in this compound also play an important role in the transport behavior. A careful look at the bandstructure (Fig.~\ref{fig: Fig. 0.}(a)) suggests that three low lying spin degenerate bands (1,2), (3,4), and (5,6), primarily of t$_{2g}$ characters formed by the distorted octahedra, cross the Fermi level at several $k$ points in the Brillouin zone. The degenerate bands (5,6) form an electron pocket centered at $\Gamma$ as shown in Fig.~\ref{fig: Fig. 0.}(g). Similarly the degenerate bands (1,2) form a hole pocket centered at R and is illustrated through Figs.~\ref{fig: Fig. 0.}(h-i). While the electron pocket forms a closed shell  as the bands crosses the Fermi level in every direction with $\Gamma$ as the center, for the case of hole pocket, an open shell is formed around $R$. This is due to the fact that along certain directions with respect to R, the bands (1,2) do not cross the Fermi level. As we raise the energy, which can be achieved by electron doping, the closing of the shell gradually occurs which is reflected in Fig.~\ref{fig: Fig. 0.}(i) where the constant energy surface is plotted for E = $\varepsilon_F + 0.5$ eV.

\subsection{\label{sec:3.1} Discussion}
We now discuss the plausible mechanisms responsible for the remarkably large quasi-quadratic magnetoresistance (MR) observed in SrTaO$_3$ thin films. As per the semiclassical two band model, for the perfect compensated semimetals, where electron and hole carriers are perfectly equal typically gives rise to large, non-saturating magnetoresistance (MR) with a quadratic magnetic-field dependence ($m \approx 2$)~\cite{PhysRevB.92.180402}. 
However, in the present case, the extracted $m$ varies from $\sim$ 1.47 at 2~K to $\sim$1.7 at 10~K [Fig.~\ref{fig: FIG. 2.}(f)], indicating a clear deviation from the ideal quadratic behavior. This suggests that charge carrier compensation is not the underlying mechanism responsible for the observed large, non-saturating MR. Insights from the calculated band structure of strained SrTaO$_3$ [Fig.~\ref{fig: Fig. 0.}(d)] reveal the coexistence of an electron pocket at the $\Gamma$ point and a hole pocket at the $R$ point, together with the two nodal loops characterized by linear crossings along the X-M direction. These two nodal loops are located approximately 0.25 eV above and 0.1 eV below the Fermi level, respectively. Any change in the carrier concentration, arising from either intrinsic or extrinsic doping, can shift the Fermi level, bringing these nodal loops closer to or across the Fermi level, thereby generating additional electron and hole pockets contributing to electron transport. Furthermore, a topologically trivial gapped near-crossing along M-$\Gamma$ (see Fig. 5 (d)) in the vicinity of the nodal line may influence electron transport. Overall, the observed electron transport arises from multiple electron and hole pockets, consistent with a multiband Hall response and the violation of conventional Kohler's scaling. It is noted that the obtained charge carrier densities and mobilities are comparable to those reported for the nodal-line semimetal ZrB$_2$~\cite{PhysRevB.97.205105} where the nodal line lies below 0.2 to 0.4 eV from the Fermi level. In addition, a similar situation has been reported in SrIrO$_3$~\cite{PhysRevB.95.121102,ren2021tailoring}, where the Dirac point lies nearly 0.3 eV below the Fermi level, yet the large non-saturating linear MR has been attributed to the presence of such linear-dispersion bands. This analogy suggests that even when the nodal or Dirac features do not lie exactly at $E_F$, their proximity can still influence the transport by providing light-mass, high-mobility carriers. We conclude that the large quasi-quadratic magnetoresistance observed in SrTaO$_3$ thin films originates from multiband transport involving conventional electron and hole pockets, together with an additional contribution from the nodal-line states.\\

\section{\label{sec:5}conclusion}
In summary, we have demonstrated that epitaxial strain of $\sim$2.6\% imposed by the SrTiO$_3$ (001) substrate significantly modifies the electronic structure and magnetotransport in SrTaO$_3$ thin films. The strained films exhibit a large, non-saturating magnetoresistance and a pronounced nonlinear Hall response, consistent with multicarrier transport and are reminiscent of behaviors widely reported in topologically non-trivial semimetals. Analysis using a semiclassical two-band model yields carrier densities and mobilities on the order of $10^{21}~\mathrm{cm}^{-3}$ and $10^{4}$\textcolor{blue}{$~\mathrm{cm}^{2}\mathrm{V}^{-1}\mathrm{s}^{-1}$}, respectively. First-principles calculations complemented with the tight binding model studies reveal that epitaxial strain leads to $a^0a^0c^+$ octahedral rotation pattern, which exhibits topological nodal-lines in the valence and conduction band edges between high-symmetry k-points X and M, which are in addition to the existence of electron and hole pockets in the system. The coexistence of topological nodal-lines with trivial electron and hole pockets in the vicinity of the Fermi level is responsible for the observed large non-saturating quasi-quadratic MR. These findings highlight the potential of strain-induced octahedral engineering in oxide heterostructures as a promising route for realizing exotic topological phases.

\begin{acknowledgments}
D.S. and A.M. acknowledge the support from the Institute of Physics, Bhubaneswar and thank S. D. Mahanti for valuable discussions.
\end{acknowledgments}

\bibliography{bib} 

\normalsize\clearpage
\begin{onecolumngrid}
	\begin{center}
		{\fontsize{12}{12}\selectfont
			\textbf{Supplementary Information: \\Strain-induced non-trivial band topology and large magnetoresistance in epitaxial SrTaO$_{3}$ thin films\\}}

            {\normalsize Aditya Mehta\,\orcidlink{0009-0003-2926-6793}$^{1,2}$\,\thanks{\href{mailto:aditya.mehta@iopb.res.in}{aditya.mehta@iopb.res.in}}, Soumyakanta Panda\, \orcidlink{0000-0001-6548-2010}$^{3}$, Pratik Sahu\,\orcidlink{0000-0003-0346-5682}$^{4,5}$, Amit Chauhan$^{6}$, Dibyajyoti Sahoo$^{7}$, Shwetha G Bhat$^{8}$, Rohit Soni$^{7}$, Goutam Tripathy$^{1,2}$, Niharika Mohapatra\,\orcidlink{0000-0002-3182-8818}$^{3}$, B. R. K. Nanda\,\orcidlink{0000-0001-8330-4824}$^{4,5}$, D. Samal\,\orcidlink{0000-0003-2618-4445}$^{1,2}$,\thanks{\href{mailto:d.samal@iopb.res.in}{dsamal@iopb.res.in}}\\

\textsuperscript{1}Institute of Physics, Bhubaneswar 751005, India\\
\textsuperscript{2}Homi Bhabha National Institute, Anushaktinagar, Mumbai 400094, India\\
\textsuperscript{3}School of Basic Sciences, Indian Institute of Technology Bhubaneswar, Bhubaneswar, Odisha 752050, India\\
\textsuperscript{4}Center for Atomistic Modelling and Materials Design, Indian Institute of Technology Madras, Chennai 600036, India\\
\textsuperscript{5}Condensed Matter Theory and Computational Lab, Department of Physics, Indian Institute of Technology Madras, Chennai 600036, India \\
\textsuperscript{6}Peter Gr\"unberg Institute, Forschungszentrum J\"ulich, 52425 J\"ulich, Germany \\
\textsuperscript{7}Department of Physics, Indian Institute of Science Education and Research Berhampur, Berhampur-760010, India \\
\textsuperscript{8}Department of Physics, Indian Institute of Science Bangalore-560012, India}
	\end{center}

\newcounter{defcounter}
	\setcounter{defcounter}{0}
	\setcounter{figure}{0}
  \renewcommand{\figurename}{Figure}
	\renewcommand{\thefigure}{S\arabic{figure}}
	\renewcommand{\tablename}{Table}
    \renewcommand{\thetable}{S\arabic{table}}
\renewcommand{\thesection}{\arabic{section}}
\setcounter{section}{0}

\vspace{0.03cm}

\subsection*{I. X-ray reflectivity}
\begin{figure*}[htbp]
\centering 
\includegraphics[scale=0.5]{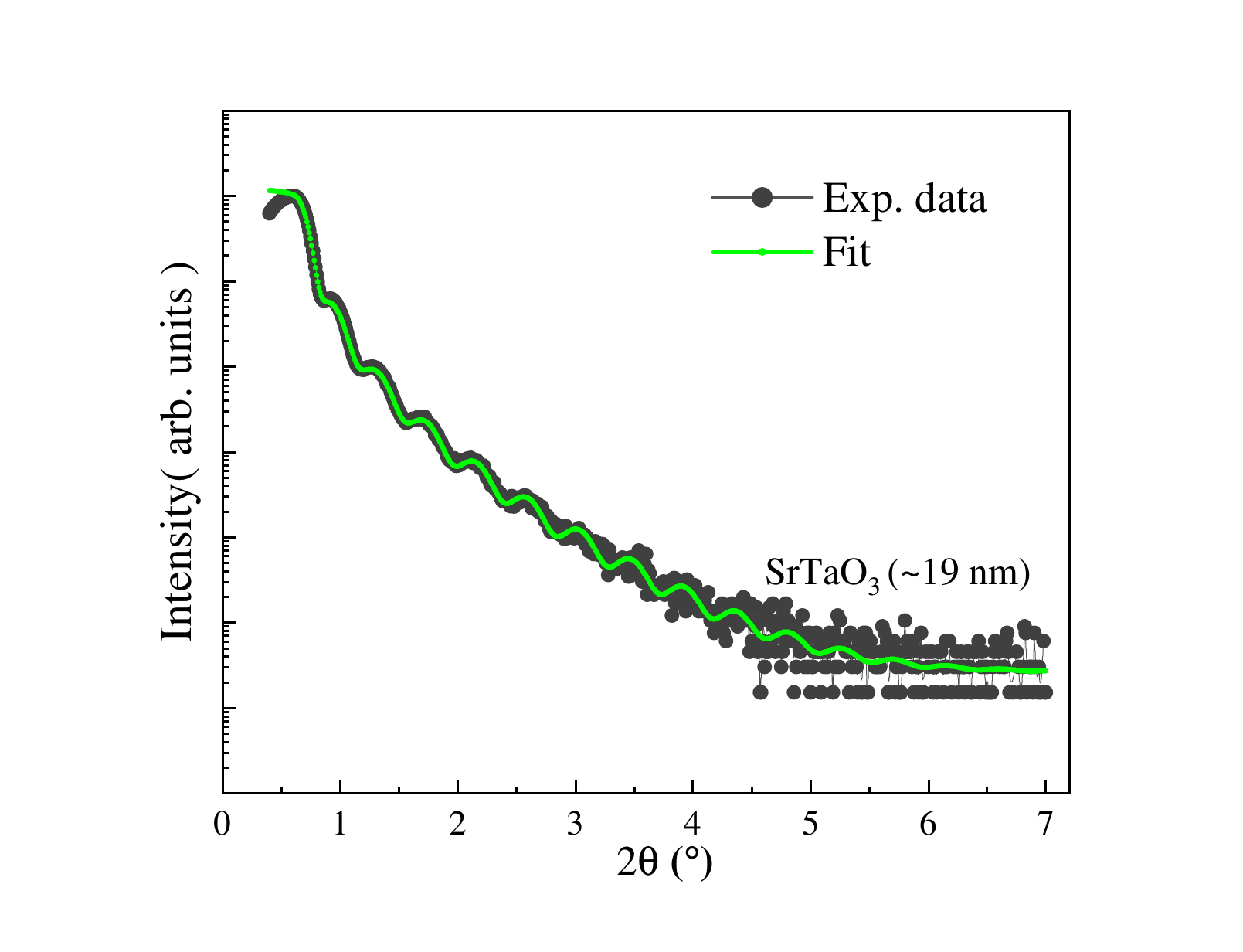}
\caption{(a) The XRR data of SrTaO$_3$ thin film, through the fit, we estimated the thickness to be 19 nm.}
\label{fig: FIG. S1.}
\end{figure*}

\clearpage 
\subsection*{II. Longitudinal and transverse conductivities}
\begin{figure*}[htbp]
\centering 
\includegraphics[scale=0.25]{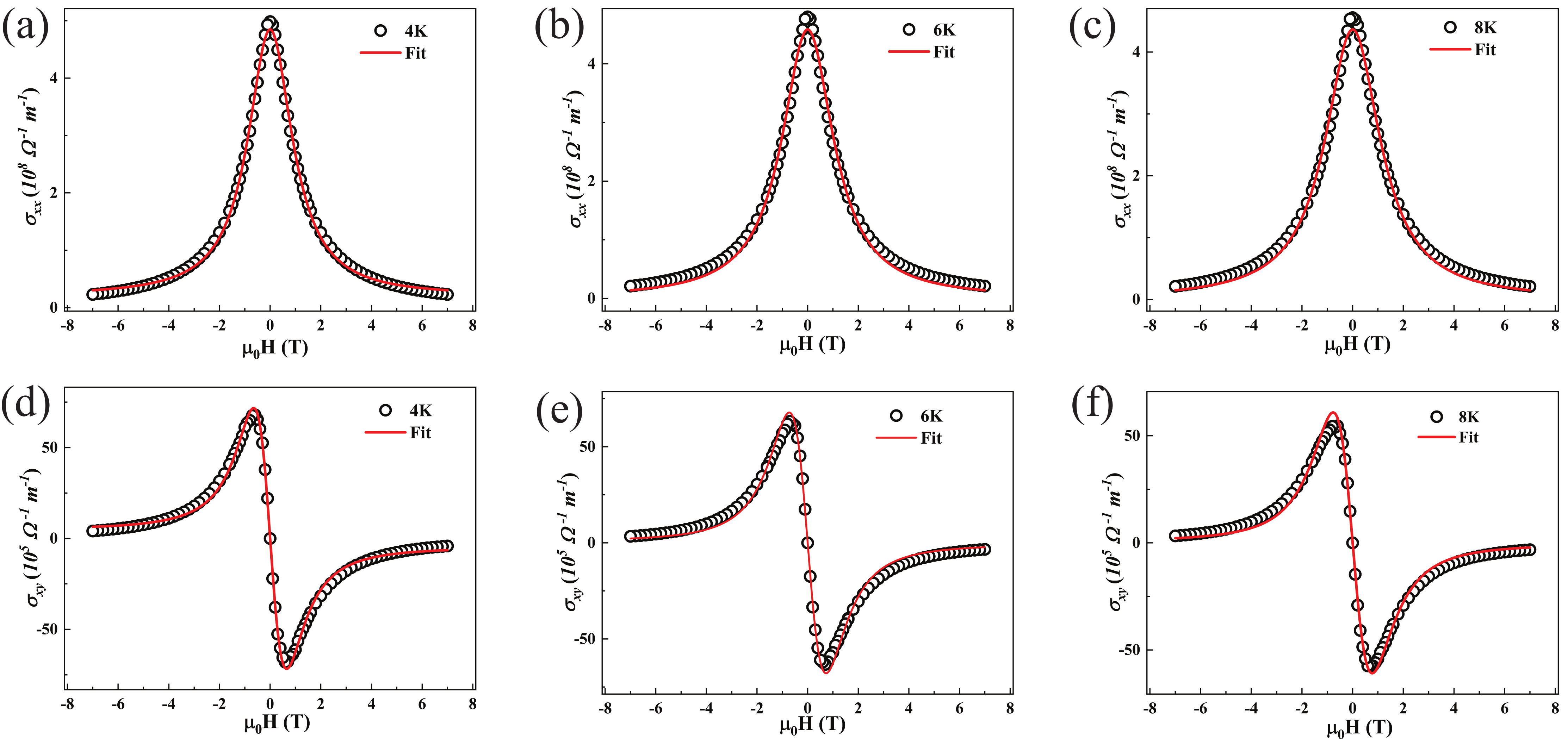}
\caption{(a-f) Simultaneous fitting of the longitudinal and transverse conductivities at 4 K, 6 K, and 8 K using a two-carrier model.}
\label{fig: FIG. S3.}
\end{figure*}
\clearpage 

\subsection*{III. Different Glazer patterns}
\begin{figure*}[htbp]
\centering 
\includegraphics[scale=0.7]{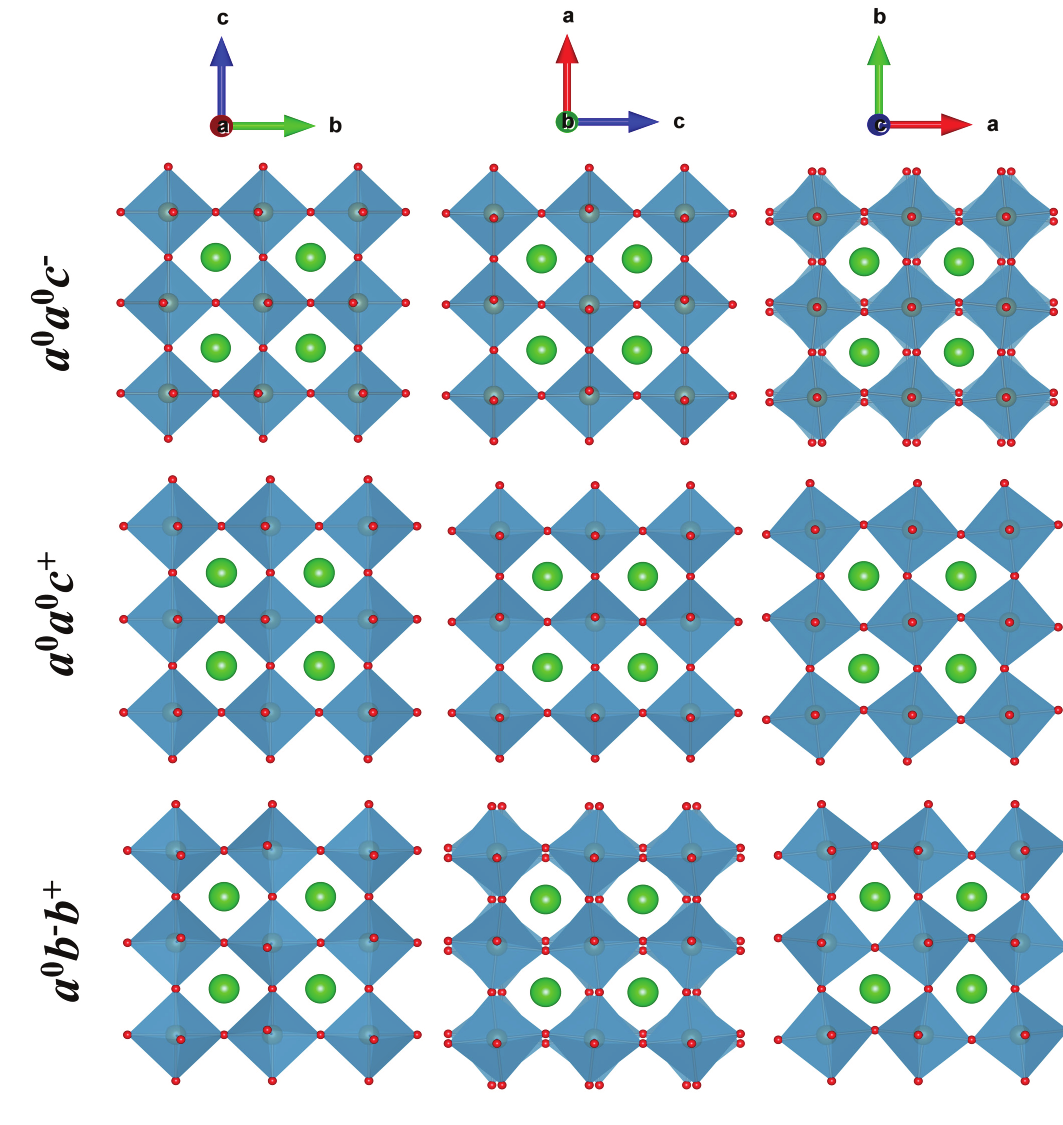}
\caption{Planar view of different Glazer patterns, namely, $a^0a^0c^-$, $a^0a^0c^+$, and $a^0b^-b^+$. The $a^0a^0c^-$ and $a^0a^0c^+$ patterns possess out-of- and in-phase rotations about $c$ axis. For $a^0b^-b^+$ pattern, there is out-of- and in-phase rotations about crystal axes $b$ and $c$.}
\label{fig: FIG. S4.}
\end{figure*}
\clearpage
\subsection*{IV. Dirac Nodal line in the $a^0a^0c^-$ configuration}
In this section, we discuss the topological properties of the $a^0a^0c^-$ phase of SrTaO$_3$, with a rotation angle $\theta = 12^0$. The unit cell for the above is twice that along the c-direction, as compared to the case of $a^0a^0c^-$. As a result, the BZ high-symmetry points remains same in the plane $k_z = 0$. The band structure shown in Fig.~\ref{fig:placeholder}(a) indicates that the four fold degeneracy remains intact even in the present case. Additionally, we computed the Berry curvature (shown in Fig.~\ref{fig:placeholder} (b)) along the same k-path, which shows large value along X-M that further validates the Dirac nodal line. Hence, we conclude that the Dirac nodal line exists in SrTaO$_3$ for both $a^0a^0c^-$ and $a^0a^0c^+$ configurations with $\theta = 12^0$.
\begin{figure}[!htb]
    \centering
    \includegraphics[width=\linewidth]{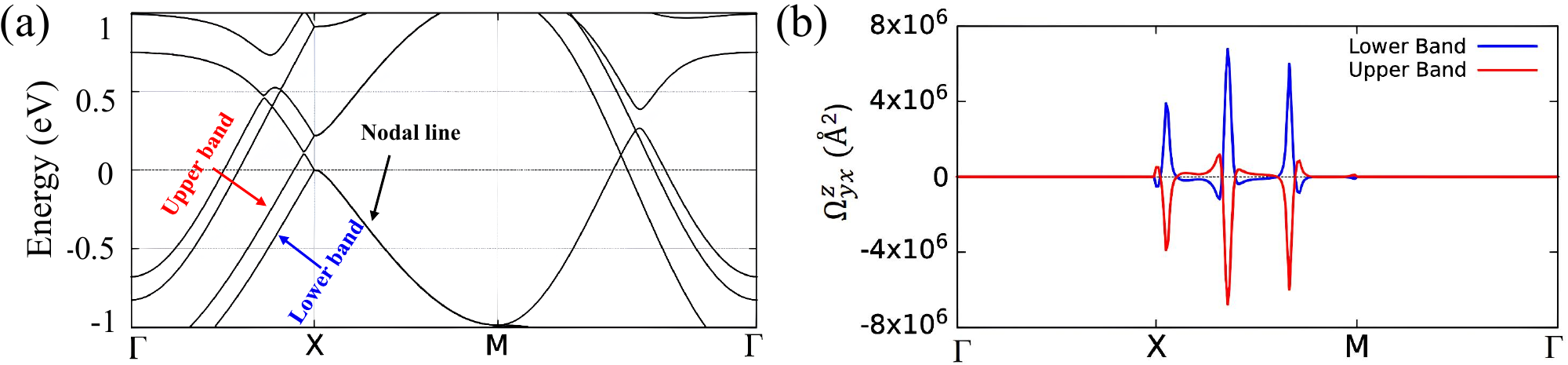}
    \caption{(a) Band structure and (b) Berry curvature component $\Omega^z_{yx}$ of SrTaO$_3$ with octahedral rotation $a^0a^0c^-$, having rotation angle $\theta = 12^0$ plotted along the k-path $\Gamma-X-M-\Gamma$.  }
    \label{fig:placeholder}
\end{figure}
\end{onecolumngrid}

\end{document}